\documentclass[12pt,a4paper]{article}
 
\usepackage{jheppub}
\usepackage[usenames,dvipsnames]{xcolor}
 
\usepackage{amssymb} 
\usepackage{amsmath}
\usepackage{mathtools}
\usepackage{amsfonts}    
\usepackage{dsfont}
\usepackage{pdfpages}
\usepackage{verbatim}
\usepackage{stmaryrd}
\usepackage{graphicx}
\usepackage{cancel}
\usepackage{tensor}
\usepackage{mathrsfs}
\usepackage{textgreek} 
\usepackage[mathscr]{euscript}
\usepackage[smalltableaux]{ytableau}
\ytableausetup{centertableaux}
\usepackage{tikz}
\usetikzlibrary{decorations.pathreplacing, decorations.markings,calc,shapes.misc,decorations.pathmorphing,patterns.meta, math}
\usepackage{slashed}
\usepackage{datetime}
\usepackage{hyperref}
\hypersetup{
    pdfencoding=unicode,
	colorlinks=true,
	urlcolor=Maroon,
	linkcolor=RoyalBlue,
	citecolor=Maroon,
	pdftitle={The Generating Function of Sunset},
	pdfauthor={Bo Feng, Liang Zhang},
	pdfdisplaydoctitle=true,
	pdfstartview=FitH,
	linktocpage=true
}

\usetikzlibrary{shapes}

\newcommand{\ba}{\begin{align}}
\newcommand{\ea}{\end{align}}
\newcommand{\be}{\begin{equation}}
\newcommand{\ee}{\end{equation}}
\newcommand{\bea}{\begin{eqnarray}}
	\newcommand{\eea}{\end{eqnarray}}

\def\bd{\begin{tikzpicture}}
\def\ed{\end{tikzpicture}}

\newcommand\Res{\mathop{\text{Res}}}
\allowdisplaybreaks

\def\XXint#1#2#3{{\setbox0=\hbox{$#1{#2#3}{\int}$}
     \vcenter{\hbox{$#2#3$}}\kern-.5\wd0}}

\definecolor{light-gray}{gray}{0.75}
\newcommand{\nn}{\nonumber \\}

\def\W #1{\widetilde{#1}}

\newcommand\Tr{\mathrm{Tr}}

\renewcommand{\geq}{\geqslant}

\title{Hidden Zeros and $2$-split via BCFW Recursion Relation}

 \author[a,b,c,d]{Bo Feng,}\emailAdd{fengbo@csrc.ac.cn}
  \author[c]{Liang Zhang,}\emailAdd{liangzhang@csrc.ac.cn}
 \author[e]{Kang Zhou}\emailAdd{zhoukang@yzu.edu.cn}
 \affiliation[a]{State Key Laboratory of Nuclear Physics and Technology, Institute of Quantum Matter, South China Normal University, Guangzhou 510006, China}
 \affiliation[b]{Guangdong Basic Research Center of Excellence for Structure and Fundamental Interactions of Matter, Guangdong Provincial Key Laboratory of Nuclear Science, Guangzhou 510006, China}
 \affiliation[c]{Beijing Computational Science Research Center, Beijing 100084, China}
 \affiliation[d]{Peng Huanwu Center for Fundamental Theory, Hefei, Anhui, 230026, China}
 \affiliation[e]{Center for Gravitation and Cosmology, College of Physical Science and Technology, Yangzhou University, No.180, Siwangting Road, Yangzhou, 225009, China}

\abstract{In this paper, we provide another angle to understand recent discoveries, i.e., the hidden zeros and corresponding 2-split behavior using the BCFW recursion relation. For the hidden zeros, we show that although the BCFW recursion relation is not directly applicable for computing amplitudes of the non-linear sigma model, we can indeed prove the zeros using the modified BCFW recursion relation. Our work also indicates that for the 2-split to hold, the current should be carefully defined. 
} 

\begin{document}

\setcounter{tocdepth}{3}
\maketitle
\setcounter{page}{2}

\section{\label{sec:intro}Introduction}

Over the past few decades, investigations into the $S$-matrix have uncovered remarkable structures in scattering amplitudes—structures that remain obscured when examined solely through traditional Lagrangian formulations and Feynman rules. Notable examples include the Bern–Carrasco–Johansson (BCJ) color-kinematics duality \cite{Bern:2008qj,Bern:2010ue,Bern:2019prr}, the Cachazo–He–Yuan (CHY) formalism \cite{Cachazo:2013hca,Cachazo:2013iea,Cachazo:2014nsa,Cachazo:2014xea}, and various geometric and combinatorial approaches \cite{Arkani-Hamed:2013jha,Arkani-Hamed:2017mur,Arkani-Hamed:2023lbd,Arkani-Hamed:2023jry}.

These developments have significantly deepened our understanding of both quantum field theory and the structure of scattering amplitudes. More recently, a novel and striking feature—termed  ``hidden zeros"—has been proposed. This structure reveals that a wide class of tree-level amplitudes vanish at special loci in kinematic space. Such behavior was first identified in ${\rm Tr}(\phi^3)$, Yang–Mills (YM), and nonlinear sigma model (NLSM) theories using techniques such as kinematic mesh constructions and stringy curve integrals \cite{Arkani-Hamed:2023swr}. 

For instance, consider a color-ordered $n$-point ${\rm Tr}(\phi^3)$ amplitude ${\cal A}^{\rm Tr(\phi^3)}_n(1,\dots,n)$. Selecting a pair of external legs $i$ and $j$, and partitioning the remaining legs into two sets—$A = \{i+1,\dots,j-1\}$ and $B = \{j+1,\dots,i-1\}$—the amplitude vanishes under the following condition:
\begin{align}
	{\cal A}^{\rm Tr(\phi^3)}_n(1,\dots,n) &\xrightarrow[]{\eqref{kine-condi-0phi}} 0, \label{0-phi}
\end{align}
at the kinematic locus defined by
\begin{align}
	s_{a, b} = 0,\quad \text{for all } a \in A,\ b \in B, \label{kine-condi-0phi}
\end{align}
where $s_{i\cdots j}$ is the usual Mandelstam variable $s_{i\cdots j}=(k_i+\cdots k_j)^2$. In the case of YM theory, additional conditions involving polarization vectors are required for the hidden zeros to emerge:
\begin{align}
	\epsilon_a \cdot \epsilon_b = \epsilon_a \cdot k_b = \epsilon_b \cdot k_a = 0. \label{eq:zeroforspin}
\end{align}
Following the pioneering work of \cite{Arkani-Hamed:2023swr}, hidden zeros have been explored from various perspectives and extended to a broader class of  theories, including Dirac–Born–Infeld (DBI), special Galileon (SG), and general relativity (GR) \cite{Rodina:2024yfc,Bartsch:2024amu,Li:2024qfp,Zhang:2024iun,Zhou:2024ddy,Zhang:2024efe}.

Meanwhile, a closely related phenomenon, known as the ``$2$-split", has also been discovered and has attracted considerable attention in recent research \cite{Cao:2024gln,Arkani-Hamed:2024fyd,Cao:2024qpp,Zhang:2024iun,Zhou:2024ddy,Zhang:2024efe}. By allowing one external leg $k$ in either set $A$ or $B$ to deviate from the hidden zeros condition—i.e., turning on $s_{k, b} \neq 0$ or $s_{k, a} \neq 0$ in Eq.~\eqref{kine-condi-0phi}—the amplitude factorizes into two amputated currents, without taking residues at any physical pole. This intriguing behavior has been found to hold for a broad class of tree-level amplitudes in both particle and string theories. Moreover, it serves as the foundation for two other important types of factorization: ``splitting near zeros" \cite{Arkani-Hamed:2023swr} and ``smooth splitting" \cite{Cachazo:2021wsz}.

To deepen our understanding of these novel properties, it will be useful to re-derive and interpret them from different perspectives. Focusing on hidden zeros and their associated $2$-split, a range of alternative derivations and interpretations have recently been proposed \cite{Rodina:2024yfc,Bartsch:2024amu,Li:2024qfp,Zhang:2024iun,Zhou:2024ddy,Zhang:2024efe,Cao:2024gln,Arkani-Hamed:2024fyd,Cao:2024qpp}. In this work, we present a new angle to understanding hidden zeros and the corresponding $2$-split, distinct from those discussed in the literature. One of the motivations stems from the following observation: for theories with a finite number of interaction terms, the full Lagrangian can be reconstructed from a finite set of low-point amplitudes. This suggests that, at least for such theories, general statements about hidden zeros and $2$-split for arbitrary numbers of external legs can be deduced purely from low-point amplitudes.

To realize this idea, the Britto–Cachazo–Feng–Witten (BCFW) on-shell recursion relation \cite{Britto:2004ap,Britto:2005fq} naturally emerges as a powerful tool. Motivated by this perspective, we utilize the BCFW recursion to derive and interpret hidden zeros and $2$-split. The BCFW method has proven particularly effective for exposing  properties in tree-level amplitudes, including on-shell constructibility \cite{Benincasa:2007xk,Arkani-Hamed:2017jhn}, BCJ relations \cite{Feng:2010my}, Kawai–Lewellen–Tye relations \cite{Bjerrum-Bohr:2010diw,Bjerrum-Bohr:2010mtb,Bjerrum-Bohr:2010kyi,Feng:2010br,Du:2011js}, and certain relations among one-loop amplitudes \cite{Feng:2011fja}.  As we will demonstrate, these advantages also benefit the study of hidden zeros and $2$-split. By examining the structure of low-point amplitudes and using recursive arguments, one can track how hidden zeros and $2$-split propagate to higher-point amplitudes—without relying on the full analytic form of the amplitudes. This feature enables a compact and elegant derivation of these phenomena and offers new insights.

We now briefly summarize the main results of this note. We re-derive the hidden zeros of tree-level amplitudes in ${\rm Tr}(\phi^3)$ theory, YM, NLSM, as well as without color-ordering amplitudes such as those in  SG, DBI, and GR theories—entirely within the framework of BCFW recursion. As emphasized, the hidden zeros in these cases are determined solely by the vanishing of certain low-point amplitudes. To handle amplitudes with poorly controlled large-$z$ behavior under BCFW shifts, we employ a modified contour integral that preserves the original deformation of external momenta. This modified integral effectively reduces the degree of divergence at large $z$ by incorporating information about amplitude zeros, and can be applied to a wide range of theories.


We also investigate the $2$-split structure of amplitudes in ${\rm Tr}(\phi^3)$, YM, GR, and NLSM. For the ${\rm Tr}(\phi^3)$ theory, the $2$-split can be rigorously established. In the cases of YM and GR amplitudes, although the overall structure supports the validity of the $2$-split, certain technical aspects—such as the precise definition of currents, which are sensitive to gauge choices—require further clarification for a fully rigorous proof. For NLSM amplitudes, a complete proof is hindered by complications arising from non-vanishing boundary terms. Nevertheless, we verify the residues at finite physical poles, which provides supporting evidence for the validity of the $2$-split in NLSM.


The remainder of this note is organized as follows. In Section~\ref{sec:hidden zeros}, we employ the BCFW recursion method to derive the hidden zeros of the aforementioned amplitudes. Subsequently, in Section~\ref{sec-split}, we prove the $2$-split structure for ${\rm Tr}(\phi^3)$, YM, and GR amplitudes, and verify that the residues of NLSM amplitudes are consistent with the $2$-split behavior. Finally, in Section~\ref{sec:con}, we conclude with a brief summary and discussion of future directions.

\section{\label{sec:hidden zeros}Hidden zeros}

\begin{figure}[ht]
	\centering
	\[\vcenter{\hbox{\includegraphics[width=0.25\linewidth]{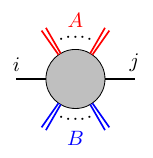}}}
    =\sum\limits_I\vcenter{\hbox{\includegraphics[width=0.25\linewidth]{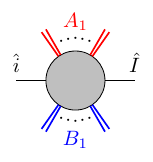}}}\frac{1}{P_I^2}\vcenter{\hbox{\includegraphics[width=0.25\linewidth]{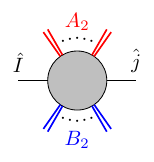}}}\]
	\caption{BCFW recursion without boundary contributions. The sum is over all possible factorization channels. Red and blue lines denote external lines in $A$ and $B$, respectively, with $A=A_1\cup A_2$ and $A_2, B=B_1\cup B_2$. Either $A_i$ or $B_j$ can be empty.}
	\label{fig:BCFW}
\end{figure}

In this section, we will prove the hidden zeros using the BCFW on-shell recursion relation. We begin by briefly reviewing the method. For a tree-level $n$-point amplitude $A_n$, we choose a pair of momenta $(k_i,k_j)$ and perform the following complex deformation:
\begin{align}
    k_i\to\widehat{k}_i(z)=k_i+z q,~ k_j\to\widehat{k}_j(z)=k_j-z q\,,
\end{align}
with  $k_i\cdot q=k_j\cdot q=q^2=0$ to preserve the momentum conservation and on-shell conditions. After the deformation, we get a rational function $A(z)$.  Let us consider the following contour integral:
\begin{align}
    \frac{1}{2\pi i}\oint_{z=0}\frac{dz}{z}A_n(z)\,,
\end{align}
where the contour is a small circle around only pole $z=0$. Evaluating the integral using Cauchy's theorem, we obtain
\begin{align}
    A_n=-\sum_{z_I}\Res_{z=z_I}\frac{A_n(z)}{z}+\Res_{z=\infty}\frac{A_n(z)}{z}=\sum_{I}\frac{A_L(z)A_R(z)}{P_I^2}+\Res_{z=\infty}\frac{A_n(z)}{z}\,.~~~\label{BCFW}
\end{align}
where $z_I$ are the finite poles of $A(z)$.
If $A_n(z)\to 0$ as $z\to \infty$, we will have $\Res_{z=\infty}\frac{A_n(z)}{z}=0$, i.e.,  the boundary contribution vanishes. We will represent this BCFW recursion relation diagrammatically in Fig.~\ref{fig:BCFW}. 

\subsection{$\Tr(\phi)^3$ amplitude}

In this subsection, we  prove the hidden zeros for color-ordered $\Tr(\phi^3)$ theory. Since
the deformed pair $(i,j)$ is non-adjacent, there is at least one propagator along the line connecting $i,j$. Consequently, ${\cal A}(z)\to 0$ as $z\to \infty$, implying the absence of boundary  contributions in \eqref{BCFW}. 
\begin{figure}[ht]
    \centering
    \includegraphics[width=0.3\linewidth]{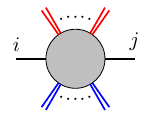}
    \caption{``zero diagram". The amplitude vanishes by imposing $s_{a,b}=0, \forall a\in A, b\in B$. For spinning particles, conditions \eqref{eq:zeroforspin} is also required.}
    \label{fig:phi3tree}
\end{figure}
We will use diagrams to present various terms in \eqref{BCFW} (see Fig.~\ref{fig:BCFW}). The red lines, corresponding to the external lines from $(i+1)$-th to $(j-1)$-th, are denoted as $A$, and the blue lines, corresponding to the external lines from $(j+1)$-th to $(i-1)$-th, are denoted as $B$. $A$ and $B$ are non-empty. There are two types of diagrams:
\begin{enumerate}
    \item[(1)] ${\cal A}_L(z)$ or/and ${\cal A}_R(z)$ contains sub-diagram which is called ``zero diagram''  as depicted in Fig.~\ref{fig:phi3tree}. Since the zero conditions can be easily verified for these zero diagrams, it follows by induction that the contribution from Fig.~\ref{fig:subtree} vanishes;
\begin{figure}[ht]
    \centering
    \includegraphics[width=0.3\linewidth]{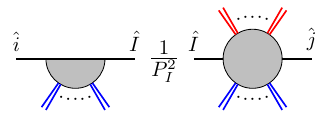}
    \hspace{0.2cm}
    \includegraphics[width=0.3\linewidth]{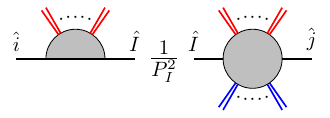}
    \hspace{0.2cm}
    \includegraphics[width=0.3\linewidth]{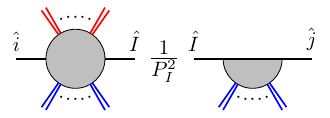}
    \hspace{0.2cm}
    \includegraphics[width=0.3\linewidth]{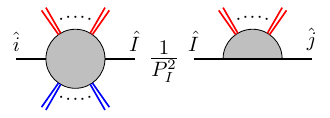}
    \hspace{0.2cm}
    \includegraphics[width=0.3\linewidth]{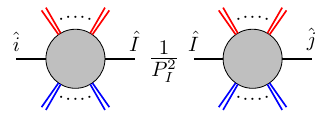}
    \caption{Since they contain ``zero diagram", their contributions vanish by induction.}
    \label{fig:subtree}
\end{figure}
    \item[(2)] Neither ${\cal A}_L(z)$ nor ${\cal A}_R(z)$ contains ``zero diagram'', as shown in Fig.~\ref{fig:boundary}. For these two diagrams, we can not use the induction.
In fact, each diagram gives nonzero contribution. 
\begin{figure}[ht]
    \centering
    \includegraphics[width=0.3\linewidth]{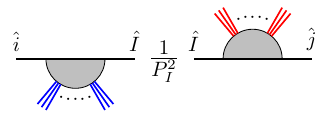}
    \hspace{1.0cm}
    \includegraphics[width=0.3\linewidth]{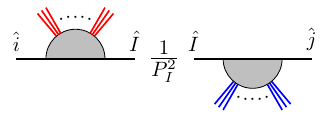}
    \caption{These diagrams cannot be ruled out by induction.}
    \label{fig:boundary}
\end{figure}
\end{enumerate}
Thus, to complete the ``hidden zeros" proof, we only need to show that the total contribution from Fig.~\ref{fig:boundary} vanish. 
 
To proceed, we first compare the pole locations  $z_I$ of the two diagrams in Fig.~\ref{fig:boundary}. For the right diagram, the on-shell condition gives
\begin{align}
    0=(\hat{k}_i+k_A)^2=(k_i+k_A+z_A q)^2=(k_i+k_A)^2+2z_A k_A\cdot q\,,
    ~~~\label{za-1}
\end{align}
where $k_A=\sum_{a\in A} k_a$. So the pole is located at
\begin{align}
    z_A=-\frac{(k_i+k_A)^2}{2k_A\cdot q}\,.\label{zA}
\end{align}
For the left diagram, the pole arises from
\begin{align}
    0=(\hat{k}_i+k_B)^2=(k_i+k_B+z_B q)^2~~\Rightarrow  z_B=-\frac{(k_i+k_B)^2}{2k_B\cdot q}\,.\label{zB}
\end{align}
Using on-shell relation, the momentum conservation and the hidden zeros condition, we get
\begin{align}\label{eq:propare}
    0=p_j^2&=(k_i+k_A+k_B)^2\notag\\
    &=k_i^2+k_A^2+k_B^2+2k_i\cdot(k_A+k_B)+{\color{red}\bcancel{2k_A\cdot k_B}}\notag\\
    &=(k_i+k_A)^2+(k_i+k_B)^2\notag\\
     \Rightarrow&~~ (k_i+k_A)^2=-(k_i+k_B)^2\,.
\end{align}
Furthermore, $k_B\cdot q=-(k_i+p_j+k_A)\cdot q=-k_A\cdot q$. That means the pole location of the two diagrams in Fig.~\ref{fig:boundary} are identical, i.e., $z_A=z_B$.
\begin{figure}[ht]
    \centering
    \includegraphics[width=0.3\linewidth]{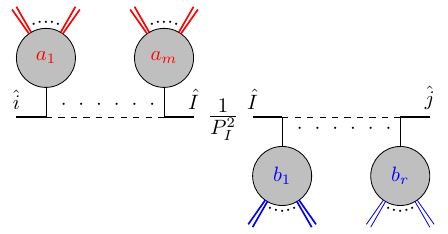}
    \hspace{0.5cm}
    \includegraphics[width=0.3\linewidth]{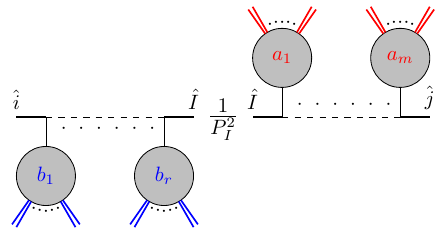}
    \caption{We partition $A$ and $B$ into smaller blocks: $A=\cup a_i, B=\cup b_j$ with $a_i\cap a_j=b_i\cap b_j=\emptyset$ and $A_k\equiv\cup_{t=1}^k a_t, B_k\equiv\cup_{t=1}^k b_t$.}
    \label{fig:partition}
\end{figure}

We now examine all possible Feynman diagrams contributing to the left diagram of Fig.~\ref{fig:boundary}. Along the line connecting to $i,j$, we partition sets $A$ and $B$ into ordered blocks, denoted as $a_i$ and $b_i$, respectively, such that $a_i\cap a_j=\emptyset$, $b_i\cap b_j=\emptyset$, $A=\cup a_i$, and $B=\cup b_i$ (see the right diagram of Fig.~\ref{fig:partition}). Furthermore, we define $A_i\equiv\cup_{t=1}^i a_t$ and $B_i\equiv\cup_{t=1}^i b_t$, thus
\begin{align}
    \emptyset\subset A_1\ldots\subset A_m\subset A,~  \emptyset\subset B_1\ldots\subset B_r\subset B\,.
\end{align}
For the right diagram, we do the corresponding division. We will prove that for the pair of the same division (as well as the same Feynman diagrams for each block), the sum of the amplitude of the two diagrams in Fig.~\ref{fig:boundary} vanishes. 

For a given partitioning, the amplitude of the right diagram in Fig.~\ref{fig:boundary} is
\begin{align}\label{eq:leftdia}
   {\cal A}^{r}&=\frac{{\cal A}_{L}^r{\cal A}_{R}^r}{(k_i+k_B)^2}\notag\\
   & =\frac{1}{\prod_k (\hat{k}_i+k_{B_k})^2}\times \frac{1}{(k_i+k_B)^2}\times \frac{1}{\prod_k (\hat{k}_i+k_{B}+k_{A_k})^2} \times {\cal S}\notag\\
& =\frac{1}{\prod_k (\hat{k}_i+k_{B_k})^2}\times \frac{1}{(k_i+k_B)^2}\times \frac{1}{\prod_k [{\color{red}\bcancel{(\hat{k}_i+k_B)^2}}+k_{A_k}^2+2k_{A_k}\cdot\hat{k}_i+{\color{red}\bcancel{2k_{A_k}\cdot k_B}}]} \times {\cal S}\notag\\
    &=\frac{1}{\prod_k (\hat{k}_i+k_{B_k})^2}\times \frac{1}{(k_i+k_B)^2}\times \frac{1}{\prod_k (k_{A_k}+\hat{k}_i)^2}\times {\cal S}\,,
\end{align}
where ${\cal S}$ represents the contributions from each block, which are identical for the pair of diagrams. To get the fourth line,  we used the on-shell relation $(\hat{k}_i+k_B)^2=0$
and zero condition $s_{a,b}=0$. Analogously, the amplitude of the left diagram is 
\begin{align}\label{eq:rightdia}
   {\cal A}^{l}&=\frac{{\cal A}_{L}^l{\cal A}_{R}^l}{(k_i+k_A)^2}\notag\\
   & =\frac{1}{\prod_k (\hat{k}_i+k_{A_k})^2}\times \frac{1}{(k_i+k_A)^2}\times \frac{1}{\prod_k (\hat{k}_i+k_{A}+k_{B_k})^2}\times {\cal S}\notag\\
    & =\frac{1}{\prod_k (\hat{k}_i+k_{A_k})^2}\times \frac{1}{(k_i+k_A)^2}\times \frac{1}{\prod_k (\hat{k}_i+k_{B_k})^2}\times {\cal S}\,.
\end{align}
Comparing \eqref{eq:leftdia} and \eqref{eq:rightdia}, it can be observed that, apart from $(k_i+k_A)^2$ and $(k_i+k_B)^2$, they are completely identical. Using the relation \eqref{eq:propare}, we conclude that ${\cal A}^l+{\cal A}^r=0$. Therefore, the sum over all factorization channels vanishes, completing the proof of the hidden zeros.

\subsection{\label{sec:YM}YM amplitudes}

For pure YM theory, there exists a choice of deformation such that the shifted amplitude satisfies ${\cal A}(z)\to 0$ as $z\to \infty$, ensuring the boundary contribution  vanishes. This implies that only the diagrams in Fig.~\ref{fig:boundary} need to be considered.. However, directly computing this contribution, as was done in the $\Tr(\phi^3)$ theory, is significantly more challenging.

Fortunately, for the YM theory, there is a crucial observation: when the deformed pair $(i,j)$ is non-adjacent, there exists a choice of $q$ such that ${\cal A}(z)\to {1\over z^2}$ when $z\to \infty$. Utilizing this property\footnote{The enhanced large $z$ behavior has been used to prove the BCJ relation \cite{Feng:2010my}. }, we can modify the BCFW recursion that cleverly avoids computing these terms in  Fig.~\ref{fig:boundary}. 

We denote $z_p\equiv-\frac{(k_i+k_A)^2}{2k_A\cdot q}=-\frac{(k_i+k_B)^2}{2k_B\cdot q}$ and consider the following contour integration
\begin{align}
    \frac{1}{2\pi i}\oint_{z=0}\frac{dz}{z}\frac{(z_p-z)\mathcal{A}(z)}{z_p}\,, \label{2.11}
\end{align}
where the contour is a small circle around only pole $z=0$. 
Since the integrand vanishes as $z\to\infty$, we can evaluate this as a sum over residues at finite poles: 
\begin{align}
    {\cal A}(0)=\frac{1}{2\pi i}\oint \frac{dz}{z}\frac{(z_p-z)\mathcal{A}(z)}{z_p}&=-\sum_I \Res\limits_{z=z_I}\left(\frac{(z_p-z){\cal A}(z)}{zz_p}\right)\notag\\
    &=\sum_I \frac{(z_I-z_p){\cal A}_L {\cal A}_R}{z_p P_I^2}\,.~~~\label{YM-BCFW}
\end{align}
The crucial point is that the added factor $(z_p-z)$ cancels  both poles in the diagrams of Fig.~\ref{fig:boundary}.  Therefore, these diagrams no longer contribute in the modified recursion \eqref{YM-BCFW}. Remaining terms in \eqref{YM-BCFW} come from Fig.~\ref{fig:subtree}. Since the modified BCFW recursion preserves the original induction structure, differing only by an overall constant factor, their contributions are zero by the same inductive argument. Thus we have proved the hidden zeros of YM theory.


\subsection{\label{sec:NLSM}NLSM amplitude}

Now we consider the hidden zeros of NLSM amplitude. It is well known that odd-point NLSM amplitudes vanish and the vertex for even multiplicity $2m$ read in the Cayley
parameterization\footnote{Hereafter, the coupling constant $\lambda$ will be absorbed by defining ${\cal A}_{2n}^{\text{NLSM}}\equiv \lambda^{2-n}A_{2n}^{\text{NLSM}}$.}
\begin{align}
    V_{2m}=-\frac{\lambda^{2m-2}}{2}\sum_{r=0}^{m-1}\sum_{i=1}^{2m}k_i\cdot k_{i+2r+1}\,.
\end{align}
Consequently, the $4$-point amplitude is ${\cal A}_{4}^{\text{NLSM}}=s_{1,2}+s_{2,3}$, and the $6$-point is
\begin{align}
    {\cal A}_{6}^{\text{NLSM}}&=s_{1,2}+s_{1,6}+s_{2,3}+s_{3,4}+s_{4,5}+s_{5,6}-\frac{(s_{1,2}+s_{2,3})(s_{4,5}+s_{5,6})}{s_{1,2,3}}\notag\\
    &-\frac{(s_{2,3}+s_{3,4})(s_{1,6}+s_{5,6})}{s_{2,3,4}}-\frac{(s_{1,2}+s_{1,6})(s_{3,4}+s_{4,5})}{s_{3,4,5}}\,.
\end{align}
From these results, it is evident that computing NLSM amplitudes via standard BCFW recursion is challenging due to their poor large-$z$ behavior and the presence of nonzero boundary contributions. In practice, they are efficiently computed using recursion relations based on soft limits \cite{Cheung:2015ota}\footnote{In \cite{Cachazo:2021wsz}, an alternative method for computing amplitudes of the NLSM is presented, leveraging information from smooth 3-splits.}. However, as we will show, the BCFW framework can still be employed to reveal the hidden zeros.

For a general $2n$-point amplitude, we perform a BCFW shift on a pair of particles $(i,j)$. The shifted NLSM amplitude can be expressed as
\begin{align}  
  {\cal A}(z)^{\text{NLSM}}= \frac{N(m+a)}{D(m)}  \,,  
\end{align}  
where $N(m+a)$ and $D(m)$ are polynomials of degree $m+a$ and $m$ in $z$, respectively. The bad large-$z$ behavior means that $a\geq 0$. This form also indicates that the shifted amplitude has $(m+a)$ zeros points, though their explicit values are generally unknown.

Following the YM case, we define $z_p\equiv-\frac{(k_i+k_A)^2}{2k_A\cdot q}=-\frac{(k_i+k_B)^2}{2k_B\cdot q}$ and consider the contour integral\footnote{Deformed contour integrations appear in various contexts, such as in \cite{Cheung:2015ota, Cachazo:2021wsz, Benincasa:2011kn, Feng:2011jxa} as well as in Eq.~\eqref{2.11}. The crucial aspect is the necessity of selecting appropriate deformations tailored to specific applications.}
\begin{align}  
    \frac{1}{2\pi i}\oint_{z=0}\frac{dz}{z}\frac{(z_p-z)\mathcal{A}(z)}{z_p\prod_{i=1}^{a+2}(z-z_i)} \,,
\end{align}  
where the contour encloses only the pole at $z=0$, and $z_i$ are the roots of $N(m+a)$. The insertion of $\prod_{i=1}^{a+2}(z-z_i)$ in denominator is to make the boundary contribution zero. Evaluating this contour integral yields 
\begin{align}  
    {\cal A}(0)&=\left(\prod_{i=1}^{a+2}(-z_i)\right)\frac{1}{2\pi i}\oint \frac{dz}{z}\frac{(z_p-z)\mathcal{A}(z)}{z_p\prod_{i=1}^{a+2}(z-z_i)}  \notag\\ 
    &=-\left(\prod_{i=1}^{a+2}(-z_i)\right)\sum_I \Res\limits_{z=z_I}\left(\frac{(z_p-z){\cal A}(z)}{zz_p\prod_{i=1}^{a+2}(z-z_i)}\right) \notag\\ 
    &=\left(\prod_{i=1}^{a+2}(-z_i)\right) \sum_I \frac{(z_I-z_p){\cal A}_L {\cal A}_R}{z_p \prod_{i=1}^{a+2}(z_I-z_i)P_I^2} \,.~~~\label{2.16}
\end{align}  
A subtle point is that the apparent poles at $(z-z_i)$ do not contribute because  $A(z_i)=0$ by construction.  Analogous to the YM amplitude, the overall constant factor does not affect the conclusion that the contributions from Fig.~\ref{fig:subtree} vanish by induction. Moreover, the factor $(z_I - z_p)$ cancels both poles in the diagrams of Fig.~\ref{fig:boundary}. Thus we have proved the hidden zeros. 

There are a few remarks regarding to above proof:
\begin{enumerate}
    \item To ensure sufficient zeros for the construction, we require $(m+a)\geq a+2$, i.e., $m\geq 2$. This implies that the method does not automatically apply to the lowest-point amplitudes, but these cases are simple enough to verify the hidden zeros property explicitly.
    \item This approach is quite general and does not require the amplitude to be color-ordered. It can also be applied to other theories, such as the SG, DBI, and GR amplitudes. The key lesson from this modified BCFW recursion relation is that, although it may not be suitable for directly computing amplitudes, it is still a powerful tool for uncovering hidden structures and properties within them.
\end{enumerate}
\section{$2$-split}
\label{sec-split}

In this section, we study the $2$-split of various theories. The general idea is similar to previous section. To prove $F=G$, we apply the same BCFW deformation to both sides, reducing the problem to proving $F(z)=G(z)$. To establish this, it suffices to show that $F(z)$ and $G(z)$ have the same pole locations (including the pole at infinity) and identical residues at each pole.

\subsection{${\rm Tr}(\phi^3)$ amplitudes}
\label{subsec-phi3-split}

%
\begin{figure}[ht]
  \centering
  \includegraphics[width=14cm]{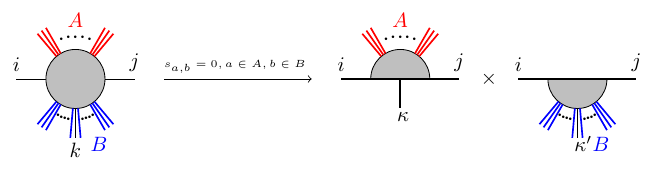}
  \caption{$2$-split of a ${\rm Tr}(\phi^3)$ amplitude.}
  \label{split}
\end{figure}

We begin with the simplest example, i.e., tree-level ${\rm Tr}(\phi^3)$ amplitudes, which exhibit the following $2$-split. For an $n$-point amplitude ${\cal A}^{{\rm Tr}(\phi^3)}_n(1,\cdots,n)$, suppose one select three external legs $i$, $j$ and $k$, with $i<j<k$ (in the sense of cyclically invariant color ordering), and separate the remaining legs into to two sets $A=\{i+1,\cdots,j-1\}$ and $B=\{j+1,\cdots,k-1\}\cup\{k+1,\cdots,i-1\}$. Then the amplitude factorizes into two amputated currents as follows \cite{Cao:2024qpp}
%
\begin{align}
{\cal A}^{{\rm Tr}(\phi^3)}_{n}(1,\cdots,n)\,\xrightarrow[]{\eqref{kine-condi-splitphi}}\,{\cal J}^{\rm{Tr}(\phi^3)}_{n_1}(i,A,j,\kappa)\,\times\,{\cal J}^{{\rm Tr}(\phi^3)}_{n+3-n_1}(j,B(\kappa'),i)\,,~~~~\label{split-phi}
\end{align}
with the loci in kinematic space
\begin{align}
k_a\cdot k_b=0\,,~~~~{\rm for}~a\in A\,,~b\in B\,,~~~~\label{kine-condi-splitphi}
\end{align}
as illustrated in Fig.~\ref{split}. For ordered amplitudes, the $\kappa$ and $\kappa'$ are inserted at the position of $k$. The notation $B(\kappa')\equiv B\cup\{\kappa'\}$ preserves the ordering\footnote{For example, $B=\{1,2,4,5\},\kappa'=3$, then $B(\kappa')=\{1,2,3,4,5\}$.}. Each resulting current carries an off-shell leg $\kappa$ or $\kappa'$, which is forced to 
be $k_\kappa=k_k+k_B$
and $k_{\kappa'}=k_k+k_A$, respectively by the momentum conservation.

We  aim to prove \eqref{split-phi} via the BCFW recursion relation.
The lowest $3$-point amplitudes trivially factorize as $1=1\times 1$, namely,
\begin{align}
{\cal A}^{{\rm Tr}(\phi^3)}_3\,\to\,{\cal A}^{{\rm Tr}(\phi^3)}_3\,\times\,{\cal A}^{{\rm Tr}(\phi^3)}_3\,.~~~\label{split-phi-3p}
\end{align}
With the above input, the general $2$-split in \eqref{split-phi} can be derived recursively using standard BCFW methods and the hidden zero conditions. With the deformation $\hat{k}_i=k_i+zq$, $\hat{k}_j=k_j-zq$ the amplitude can be written as 
\begin{align}
{\cal A}^{{\rm Tr}(\phi^3)}(1,\cdots,n)=\sum_I\,{{\cal A}^{{\rm Tr}(\phi^3)}_L(z_I)\,{\cal A}^{{\rm Tr}(\phi^3)}_R(z_I)\over P_I^2}\,,~~~\label{3.4}
\end{align}
where we have used the fact that since $i,j$ are non-adjacent, the large $z$ behavior is $z^{-1}$ and hence the residue at $z=\infty$ vanishes. 

\begin{figure}[ht]
  \centering
  \includegraphics[width=0.4\linewidth]{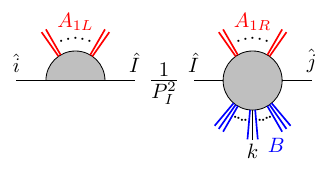}
   \includegraphics[width=0.4\linewidth]{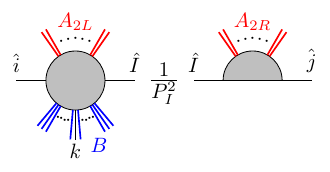}
   \includegraphics[width=0.4\linewidth]{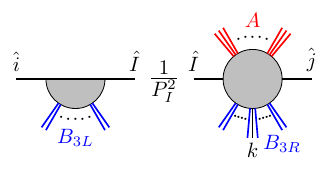}
   \includegraphics[width=0.4\linewidth]{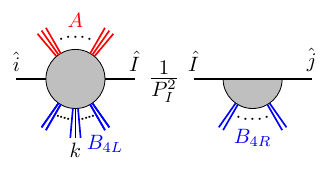}
  \caption{Non-vanishing contributions to the amplitude subject to the 2-split constraints.}
  \label{red-blue}
\end{figure}

Among all terms in BCFW recursion relation, some vanish due to the hidden zeros. The only nonzero contributions come from the following four sectors:
\begin{align}
\label{residue-splitphi}
{\cal A}^{{\rm Tr}(\phi^3)}(1,\cdots,n)
=&\sum_{A_{1R}}\,{\cal A}^{\rm{Tr}(\phi^3)}(\hat{i},A_{1L},\hat{I})\,{1\over P_I^2}\,{\cal A}^{\phi^3}(\hat{I},A_{1R},j,B(k))\nn
& + \sum_{A_{2L}}\,{\cal A}^{\rm{Tr}(\phi^3)}(B(k),\hat{i},A_{2L},\hat{I})\,{1\over P_I^2}\,{\cal A}^{\phi^3}(\hat{I},A_{2R},\hat{j})\nn
& + \sum_{B_{3R}}\,{\cal A}^{\rm{Tr}(\phi^3)}(B_{3L},\hat{i},\hat{I})\,{1\over P_I^2}\,{\cal A}^{\phi^3}(\hat{I},A,\hat{j},B_{3R}(k))\nn
& + \sum_{B_{4L}}\, {\cal A}^{\rm{Tr}(\phi^3)}(B_{4L}(k),\hat{i},A,\hat{I})\,{1\over P_I^2}\,{\cal A}^{\phi^3}(\hat{j},B_{4R},\hat{I})\,,
\end{align}
as  diagrammatically illustrated in Fig.~\ref{red-blue}.

\begin{figure}[ht]
  \centering
  \includegraphics[width=0.8\linewidth]{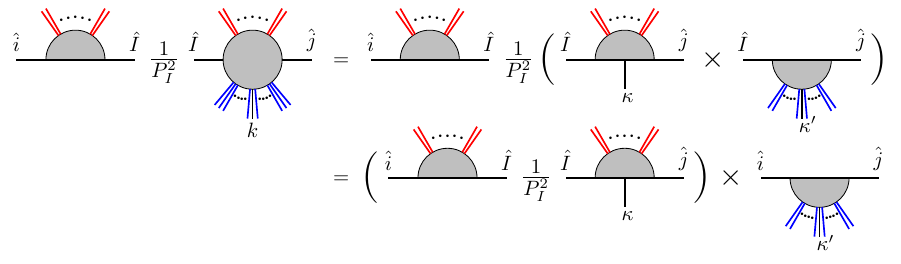}
  \caption{Reducing the residue of amplitude to the residue of a current.}
  \label{resequ}
\end{figure}

For each term in \eqref{residue-splitphi}, we apply the split result given in \eqref{split-phi} to the sub-amplitude containing the leg $k$. To clarify the discussion, let us focus on the first line of \eqref{residue-splitphi}. The on-shell amplitude  ${\cal A}^{{\rm Tr}(\phi^3)}(\hat{I},A_{1R},j,B(k))$ factorizes as
\begin{align}
{\cal A}^{{\rm Tr}(\phi^3)}(\hat{I},A_{1R},j,B(k))\xrightarrow[]{\eqref{kine-condi-splitphi}}\,{\cal J}^{\rm{Tr}(\phi^3)}(\hat{I},A_{1R},\hat{j},\kappa_I)\,\times\,{\cal J}^{{\rm Tr}(\phi^3)}(\hat{j},B(\kappa_I'),\hat{I})\,,
\end{align}
and this term is turned to
\begin{align}\label{resA-resJ}
& {\cal A}^{\rm{Tr}(\phi^3)}(\hat{i},A_{1L},\hat{I})\,{1\over P_I^2}\,{\cal A}^{{\rm Tr}(\phi^3)}(\hat{I},A_{1R},j,B(k))\nn
=&\Big({\cal A}^{\rm{Tr}(\phi^3)}(\hat{i},A_{1L},\hat{I})\,{1\over P_I^2}\,{\cal J}^{\rm{Tr}(\phi^3)}(\hat{I},A_{1R},\hat{j},\kappa_I)\Big)\,\times\,{\cal J}^{{\rm Tr}(\phi^3)}(\hat{j},B(\kappa_I'),\hat{I})\nn
=&\left(-{\rm Res}\big|_{z=z_I}\,{{\cal J}_{n_1}^{\rm{Tr}(\phi^3)}(\hat{i},A,\hat{j},\kappa)\over z}\right)\,\times\,{\cal J}_{n+3-n_1}^{{\rm Tr}(\phi^3)}(\hat{j},B(\kappa_I'),\hat{I})\nn
=& \left(-{\rm Res}\big|_{z=z_I}\,{{\cal J}_{n_1}^{\rm{Tr}(\phi^3)}(\hat{i},A,\hat{j},\kappa)\over z}\right)\,\times\,{\cal J}_{n+3-n_1}^{{\rm Tr}(\phi^3)}(\hat{j},B(\kappa'),\hat{i})\,.
\end{align}
In the second line,  momentum conservation leads to the identities
\bea 
k_{\kappa_I'} =  k_{k}+k_{A_{1R}},~~~~ k_{\kappa_I} =  k_k+k_B=k_{\kappa}\,.
\eea
From the third line to the fourth line, we have used a key observation:
\begin{align}
\label{key-observation}
{\cal J}_{n+3-n_1}^{{\rm Tr}(\phi^3)}(\hat{j},B(\kappa'_I),\hat{I})={\cal J}_{n+3-n_1}^{{\rm Tr}(\phi^3)}(\hat{j},B(\kappa'),\hat{i})\,,
\end{align}
which holds for the following reasons. First, consider the propagators in the left current along the line connecting $\hat{I}$ and $\hat{j}$ which take the form 
${1\over \prod_{B'} (\hat{k}_I+k_{B'})^2}\times {1\over \prod_{B''} (\hat{k}_j+k_{B''})^2}$, where $B',B''$ are  arbitrary subsets of $B$, and $k_{B'}, k_{B''}$ denote the total momenta of the respective $B',B''$ blocks. A straightforward computation yields
\bea 
 \label{observe1}
(\hat{k}_I+k_{B'})^2 &= & k_{B'}^2+2\hat{k}_I\cdot k_{B'}=k_{B'}^2+2(\hat{k}_i+k_{A_{1L}})\cdot k_{B'}\nn
& = & k_{B'}^2+2\hat{k}_i\cdot k_{B'}
=(\hat{k}_i+k_{B'})^2\,,
\eea
where we used the condition \eqref{kine-condi-splitphi} and the fact that $\hat{k}_i^2=\hat{k}_I=0$. Secondly, for the block containing the leg $\kappa_I'$ in the current ${\cal J}_{n+3-n_1}^{{\rm Tr}(\phi^3)}(\hat{j},B(\kappa_I'),\hat{I})$, the inverse of massive propagators is given by 
 \bea 
 (k_{\kappa_{I}'}+k_{B_i})^2- (k_{\kappa_{I}'})^2 &=& k_{B_i}^2+2k_{B_i}\cdot k_{\kappa_{I}'} = k_{B_i}^2+2k_{B_i}\cdot (k_k+k_{A_{1R}}) \nn
 &=& k_{B_i}^2+2k_{B_i}\cdot (k_k+k_{A}) = (k_{\kappa'}+k_{B_i})^2- (k_{\kappa'})^2\,.
 \eea
The above manipulations are illustrated in Fig.~\ref{resequ}. 

A similar treatment applies to the remaining three lines in \eqref{residue-splitphi}. Consequently, the right-hand side  of \eqref{residue-splitphi} can be rewritten as 
\begin{align}\label{resA-resJ-2}
	& \textstyle -\sum_{I_1}{\rm Res}\big|_{z=z_{I_1}}\left({{\cal J}^{{\rm Tr}(\phi^3)}_{n_1}(z)\over z}\right)\times
{\cal J}^{{\rm Tr}(\phi^3)}_{n+3-n_1}(z_{I_1})-\sum_{I_2}{\cal J}^{{\rm Tr}(\phi^3)}_{n_1}(z_{I_2})\times{\rm Res}\big|_{z_{I_2}}\left({{\cal J}^{{\rm Tr}(\phi^3)}_{n+3-n_1}(z)\over z}\right)\nn
=&-\sum_I\,{\rm Res}\big|_{z=z_I}\,\Big({{\cal J}^{{\rm Tr}(\phi^3)}_{n_1}(z)\,\times\,{\cal J}^{{\rm Tr}(\phi^3)}_{n+3-n_1}(z)\over z}\Big)\,,
\end{align}
where the second equality holds as long as two currents do not share any overlapped pole, due to the definitions of them. The result can be written as 
\begin{align}
{\cal A}^{{\rm Tr}(\phi^3)}_n=-\sum_I\,{\rm Res}\big|_{z=z_I}\,{{\cal A}^{{\rm Tr}(\phi^3)}_n(z)\over z}
=-\sum_I\,{\rm Res}\big|_{z=z_I}\,{{\cal B}^{{\rm Tr}(\phi^3)}_n(z)\over z}\,,~~\label{res-ident-phi}
\end{align}
with the definition
\begin{align}
{\cal B}^{{\rm Tr}(\phi^3)}_n(z)={\cal J}^{{\rm Tr}(\phi^3)}_{n_1}(z)\,\times\,{\cal J}^{{\rm Tr}(\phi^3)}_{n+3-n_1}(z)\,.
\end{align}
We now consider the contour integration $\oint_{z=0} dz {{\cal B}^{{\rm Tr}(\phi^3)}_n(z)\over z}$. 
First, ${\cal B}^{{\rm Tr}(\phi^3)}_n(z)$ does not have any boundary contribution. This is because the first current ${\cal J}^{{\rm Tr}(\phi^3)}_{n_1}$, where the deformed legs $i$ and $j$ are non-adjacent in the color ordering, has at least one propagator depends on $z$, resulting in a large-$z$ behavior of $z^{-1}$. The second current ${\cal J}^{{\rm Tr}(\phi^3)_{n+3-n_1}}$, where two deformed legs are adjacent, behaves as $z^0$ at infinity. Therefore, the product ${\cal B}^{{\rm Tr}(\phi^3)}_n(z)\big|_{z\to\infty}\sim z^{-1}$.
Using this, we obtain 
\bea {\cal B}^{{\rm Tr}(\phi^3)}_n(z=0)=-\sum_{I'}\,{\rm Res}\big|_{z=z_{I'}}\,{{\cal B}^{{\rm Tr}(\phi^3)}_n(z)\over z}\,.~~\label{res-ident-phi-1}\eea
Now another observation is that the set of finite poles in \eqref{residue-splitphi} is exact the complete set of poles of ${\cal B}^{{\rm Tr}(\phi^3)}_n(z)$, based on the definitions of two currents. Combining \eqref{res-ident-phi} and \eqref{res-ident-phi-1}  we conclude that ${\cal A}^{{\rm Tr}(\phi^3)}_n={\cal B}^{{\rm Tr}(\phi^3)}_n$, which is  the $2$-split in \eqref{split-phi}.

Before ending this subsection, we briefly discuss the role of hidden zeros in the proof of \eqref{split-phi}. The recursive proof relies on three key observations. First, the lowest-point amplitudes trivially satisfy the $2$-split. Second, each residue of the deformed amplitude can be identified with the residue of a product of two deformed currents. Third, both ${\cal A}^{{\rm Tr}(\phi^3)}_n(z)$ and ${\cal B}^{{\rm Tr}(\phi^3)}_n(z)$ vanish at the infinity, or more accurately, their residues at the pole of infinity are same. 
Among these three observations,  the second one crucially depends on the hidden zeros condition. Comparing the loci \eqref{kine-condi-splitphi} in the kinematic space with the loci \eqref{kine-condi-0phi} in the previous section, we see that the condition for $2$-split arises from the hidden zeros condition by relaxing it—specifically, by allowing $s_{a,k}\neq0$ for $k\in\{j+1,\cdots,i-1\}$ (or allowing $s_{k,b}\neq0$ for $k\in\{i+1,\cdots,j-1\}$). Using the BCFW recursion relation offers an alternative perspective on how the relaxation of the hidden zeros condition gives rise to the appearance of $2$-split behavior.

\subsection{YM amplitudes}
\label{subsec-YM-split}

The $n$-point tree-level YM amplitude has a similar $2$-split \cite{Cao:2024qpp}
\begin{align}\label{split-YM}
{\cal A}^{\rm YM}_{n}(1,\cdots,n)\,\xrightarrow[\eqref{kine-condi-splitYM1}\eqref{kine-condi-splitYM2}]{\eqref{kine-condi-splitphi}}\,\epsilon_{k}\cdot{\cal J}^{\rm YM}_{n_1}(i,A,j,\kappa)\,\times\,{\cal J}^{{\rm YM}\oplus{\rm Tr}(\phi^3)}_{n+3-n_1}(j_\phi,B(\kappa'_\phi),i_\phi)\,,
\end{align} 
with additional constraints on polarizations
\begin{align}
\epsilon_a\cdot k_b=\epsilon_b\cdot k_a=\epsilon_a\cdot \epsilon_b=0\,,~~~~{\rm for}~a\in A\,,~b\in B\,,~~\label{kine-condi-splitYM1}
\end{align}
and
\begin{align}
\epsilon_c\cdot k_b=\epsilon_c\cdot \epsilon_b=0\,,~~~~{\rm with}~c=i,j,k\,.~~\label{kine-condi-splitYM2}
\end{align}
In \eqref{split-YM}, $\big({\cal J}^{\rm YM}_{n_1}(i,A,j,\kappa)\big)^\mu$ is a vector current carrying an off-shell leg $\kappa$, with all particles being gluons. On the other hand, ${\cal J}^{{\rm YM}\oplus{\rm Tr}(\phi^3)}_{n+3-n_1}(j_\phi,B(\kappa'_\phi),i_\phi)$ is a scalar current, particles $i$, $j$ and $\kappa'$ are scalars of ${\rm Tr}(\phi^3)$ theory, while the remaining particles are gluons. If we replace subscripts $b$ with $a$ in \eqref{kine-condi-splitYM2}, namely,
\begin{align}
\epsilon_c\cdot k_a=\epsilon_c\cdot \epsilon_a=0\,,~~~~{\rm with}~c=i,j,k\,,~~\label{kine-condi-splitYM3}
\end{align}
the $2$-split \eqref{split-YM} is transformed into
\begin{align}
{\cal A}^{\rm YM}_{n}(1,\cdots,n)\,\xrightarrow[\eqref{kine-condi-splitYM1}\eqref{kine-condi-splitYM3}]{\eqref{kine-condi-splitphi}}\,{\cal J}^{{\rm YM}\oplus{\rm Tr}(\phi^3)}_{n_1}(i_\phi,A,j_\phi,\kappa_\phi)\,\times\,\epsilon_k\cdot{\cal J}^{\rm YM}_{n+3-n_1}(j,B(\kappa'),i)\,.\label{split-YM-2}
\end{align}

Both $2$-split \eqref{split-YM} and \eqref{split-YM-2} can be proven using similar recursive techniques. The lowest $3$-point amplitude trivially factorizes as ${\cal A}^{\rm YM}_3={\cal A}^{\rm YM}_3\times{\cal A}^{{\rm Tr}(\phi^3)}_3$, since ${\cal A}^{{\rm Tr}(\phi^3)}_3$ is a constant that can be normalized to $1$. With this simple starting point, the general $2$-split can be established recursively.

We deform $k_i$ and $k_j$ obtaining an analogue of \eqref{residue-splitphi},
\begin{align}
\label{residue-splitYM}
   {\cal A}^{\rm YM}_{n}(1,\cdots,n) =&\sum_{A_{1R}}\sum_H\,{\cal A}^{\rm{Tr}(\phi^3)}(\hat{i},A_{1L},\hat{I})\,{1\over P_I^2}\,{\cal A}^{\phi^3}(\hat{I},A_{1R},j,B(k))\nn
& +\sum_{A_{2L}}\sum_H\, {\cal A}^{\rm{Tr}(\phi^3)}(B(k),\hat{i},A_{2L},\hat{I})\,{1\over P_I^2}\,{\cal A}^{\phi^3}(\hat{I},A_{2R},\hat{j})\nn
& +\sum_{B_{3R}}\sum_H\, {\cal A}^{\rm{Tr}(\phi^3)}(B_{3L},\hat{i},\hat{I})\,{1\over P_I^2}\,{\cal A}^{\phi^3}(\hat{I},A,\hat{j},B_{3R}(k))\nn
& +\sum_{B_{4L}}\sum_H\,  {\cal A}^{\rm{Tr}(\phi^3)}(B_{4L}(k),\hat{i},A,\hat{I})\,{1\over P_I^2}\,{\cal A}^{\phi^3}(\hat{j},B_{4R},\hat{I})\,,
\end{align}
which is again caused by hidden zeros and the absence of boundary contributions, as $A(z)\to z^{-2}$ when $i,j$ are non-adjacent, as shown in \cite{Arkani-Hamed:2008bsc}. Here $\sum_H$ denotes a sum over all possible helicities of the internal gluon $I$. Along the line in the previous subsection, the next step is to reinterpret the residues in \eqref{residue-splitYM} as residues of a product of two deformed currents. To achieve this goal, we consider the first line of \eqref{residue-splitYM}, and discuss two cases corresponding to the conditions \eqref{kine-condi-splitYM2} and \eqref{kine-condi-splitYM3}.

Let us start from the first case \eqref{kine-condi-splitYM2}. From the preceding subsection, the first step is to use the split form for the ${\cal A}^{\phi^3}(\hat{I},A_{1R},j,B(k))$. However, comparing to the scalar case, a complication arises, i.e., particle $\hat{I}$ has helicity $H$. To facilitate the inductive argument, it is necessary to demonstrate the satisfaction of condition \eqref{kine-condi-splitYM2}. While a rigorous proof of this fact is not provided, the following reasoning offers supporting evidence. Observing that the kinematic conditions \eqref{kine-condi-splitphi} and \eqref{kine-condi-splitYM1} can be understood through the division of the full kinematic space into two orthogonal complementary subspaces, namely ${\cal S}={\cal S}_1\oplus{\cal S}_2$. Within this framework, momenta and polarizations carried by legs in $A$ lie in ${\cal S}_1$, while those carried by legs in $B$ lie in ${\cal S}_2$, and the condition \eqref{kine-condi-splitYM2} means polarizations $\epsilon_c$ with $c=i,j,k$ also lie in ${\cal S}_1$. Assuming the polarization $\epsilon_I^H$ carried by ${\cal A}^{\rm YM}(\hat{i},A_{1L},\hat{I})$ lie in ${\cal S}_2$, it annihilates the amplitude, since the momenta and polarizations of the remaining gluons, excluding $\hat{k}_i$, lie in ${\cal S}_1$. For $\hat{k}_i$, we can use momentum conservation, $\hat{k}_i=-\widehat{P}_I-k_{A_{1L}}$, along with $\epsilon_I^H\cdot \widehat{P}_I=0$, to confirm $\epsilon^H_I\cdot\hat{k}_i=0$. Thus, the effective polarization $\epsilon_I^H$, as well as the polarization $\bar{\epsilon}_I^H$ carried by ${\cal A}^{\rm YM}(\hat{I}, A_{1R},\hat{j},B(k))$, must lie in ${\cal S}_1$. 

With above arguments, ${\cal A}^{\rm YM}(\hat{I}, A_{1R},\hat{j},B(k))$ factorizes as in \eqref{split-YM},
\begin{align}
{\cal A}^{\rm YM}(\hat{I}, A_{1R},\hat{j},B(k))\xrightarrow[\eqref{kine-condi-splitYM1}\eqref{kine-condi-splitYM2}]
{\eqref{kine-condi-splitphi}}\,\epsilon_k\cdot{\cal J}^{\rm YM}(\hat{I},A_{1R},\hat{j},\kappa)\,\times\,{\cal J}^{{\rm YM}\oplus{\rm Tr}(\phi^3)}(\hat{j}_\phi,B(\kappa'_{I,\phi}),\hat{I}_\phi)\,,
\end{align}
according to the spirit of recursion. Subsequently, we can merge ${\cal A}^{\rm YM}(\hat{i},A_{1L},\hat{I})$ and $\epsilon_k\cdot{\cal J}^{\rm YM}(\hat{I},A_{1R},\hat{j},\kappa)$ to get the analogue of \eqref{resA-resJ},
\begin{align}
&{\cal A}^{\rm YM}(\hat{i},A_{1L},\hat{I})\,{1\over P_I^2}\,{\cal A}^{\rm YM}(\hat{I},A_{1R},\hat{j},B(k))\nn
=&-\left({\rm Res}\big|_{z=z_I}\,{\epsilon_k\cdot{\cal J}_{n_1}^{\rm YM}(\hat{i},A,\hat{j},\kappa)\over z}\right)\,\times\,{\cal J}_{n+3-n_1}^{{\rm YM}\oplus{\rm Tr}(\phi^3)}(\hat{j}_\phi,B(\kappa'_{I,\phi}),\hat{I}_\phi)\nn
=&-\left({\rm Res}\big|_{z=z_I}\,{\epsilon_k\cdot{\cal J}_{n_1}^{\rm YM}(\hat{i},A,\hat{j},\kappa)\over z}\right)\,\times\,{\cal J}_{n+3-n_1}^{{\rm YM}\oplus{\rm Tr}(\phi^3)}(\hat{j}_\phi,B(\kappa'_\phi),\hat{i}_\phi)\,.
\end{align}
In the above equation, when from the second line to the third line, we have used a key relation
\begin{align}
	{\cal J}_{n+3-n_1}^{{\rm YM}\oplus{\rm Tr}(\phi^3)}(\hat{j}_\phi,B(\kappa'_{I,\phi}),\hat{I}_\phi)
	={\cal J}_{n+3-n_1}^{{\rm YM}\oplus{\rm Tr}(\phi^3)}(\hat{j}_\phi,B(\kappa'_{\phi}),\hat{i}_\phi)\,.\label{YM-ide}
\end{align}
which is similar to the relation \eqref{key-observation}. To prove this, we need to show the identity of 
each diagram on both sides. For the denominator, which is the product of propagators, we can do the exact same
argument as in the previous subsection \ref{subsec-phi3-split}. However, for the nontrivial numerator, the analysis is very nontrivial.

Let us first have a general analysis. Using momentum conservation, we see the numerator is the function of Lorentz contractions among variables $\{ k_b,\epsilon_b,\hat{k}_j, \hat{k}_{I_\phi}\}$. It is easy to see that 
\bea 
\hat{k}_{I_\phi}\cdot \{ k_b,\epsilon_b \}=(\hat{k}_{i}+k_{A_{1R}})\cdot \{ k_b,\epsilon_b \}=
\hat{k}_i\cdot \{ k_b,\epsilon_b \}\,,
\eea
using the condition \eqref{kine-condi-splitYM1}. The trouble part is the contraction $\hat{k}_{I_\phi}\cdot \hat{k}_j\neq \hat{k}_i\cdot \hat{k}_j$. Consequently, for
${\cal J}_{n+3-n_1}^{{\rm YM}\oplus{\rm Tr}(\phi^3)}(\hat{j}_\phi,B(\kappa'_{I,\phi}),\hat{I}_\phi)={\cal J}_{n+3-n_1}^{{\rm YM}\oplus{\rm Tr}(\phi^3)}(\hat{j}_\phi,B(\kappa'_\phi),\hat{i}_\phi) $, we must demonstrate the absence of the contraction  $\hat{k}_{I_\phi}\cdot \hat{k}_j$. 
For current case, the crucial observation is that the cubic vertex of two scalars and a gluon takes the form $\eta_{\mu\nu}(k_{1\phi}-k_{2\phi})^\mu$. Therefore, for tree diagrams with only three scalars\footnote{In the presence of four scalars, diagrams exist where scalar momenta contract with each other.}, scalar momenta can only contract with the momenta and polarization vectors of gluons.  Combining all these arguments  we have proved \eqref{YM-ide}.

For the second case, we observe that the polarization $\hat\epsilon_i$ can only contract with only $\epsilon_I^h$; otherwise the amplitude
${\cal A}^{\rm YM}(\hat{i},A_{1L},\hat{I})$ is annihilated due to the condition \eqref{kine-condi-splitYM3}. Using
\begin{align}
\hat\epsilon_{i\mu}\sum_H\,\big(\epsilon_I^{H\mu}\,\bar{\epsilon}_I^{H\nu}\big)\,\sim\,\hat\epsilon_{i\mu} g^{\mu\nu}=\hat\epsilon_{i}^\nu\,,~~\label{polar-metri}
\end{align}
the polarization $\epsilon_i$ can be absorbed into ${\cal A}^{\rm YM}(\hat{I},A_{1R},\hat{j},B_{k})$, by replacing $\bar{\epsilon}_I^H$ with $\epsilon_i$.
After this manipulation, the summation over helicities is eliminated, and ${\cal A}^{\rm YM}(\hat{i},A_{1L},\hat{I})$ transforms into a new amplitude that external particles $\hat{i}$ and $\hat{I}$ behave as scalars. Indeed, for the amplitude ${\cal A}^{\rm YM}(\hat{i},A_{1L},\hat{I})$, the effect of above procedure is equivalent to setting $\epsilon_i$ and $\epsilon_I$ to reside in extra dimensions satisfying $\epsilon_i\cdot\epsilon_I=1$. As well known, the resulting object of such dimensional reduction is the ${\rm YM}\oplus{\rm Tr}(\phi^3)$ amplitude ${\cal A}^{{\rm YM}\oplus{\rm Tr}(\phi^3)}(\hat{i}_\phi,A_{1L},\hat{I}_\phi)$ which involves two scalars $\hat{i}_\phi$ and $\hat{I}_\phi$. Thus, we arrive at
\begin{align}
&\sum_H\,{\cal A}^{\rm YM}(\hat{i},A_{1L},\hat{I})\,{1\over P_I^2}\,{\cal A}^{\rm YM}(\hat{I},A_{1R},\hat{j},B(k))\nn
=&{\cal A}^{{\rm YM}\oplus{\rm Tr}(\phi^3)}(\hat{i}_\phi,A_{1L},\hat{I}_\phi)\,{1\over P_I^2}\,{\cal A}^{\rm YM}(\hat{I}^{\epsilon_i},A_{1R},\hat{j},B(k))\,,
\end{align}
where the superscript $\epsilon_i$ is introduced to emphasize that the polarization of the particle $\hat{I}$ is $\epsilon_i$. The amplitude
${\cal A}^{\rm YM}(\hat{I}^{\epsilon_i},A_{2R},\hat{j},B(k))$ factorizes as in \eqref{split-YM-2},
\begin{align}
&{\cal A}^{\rm YM}(\hat{I}^{\epsilon_i},A_{1R},\hat{j},B(k))\nn
&\xrightarrow[]{\eqref{kine-condi-splitphi}\eqref{kine-condi-splitYM1}\eqref{kine-condi-splitYM3}}\,{\cal J}^{{\rm YM}\oplus{\rm Tr}(\phi^3)} (\hat{I}_\phi,A_{1R},\hat{j}_\phi,\kappa_{\phi})\,\times\,\epsilon_k\cdot{\cal J}^{\rm YM}(\hat{j},B(\kappa'_I),\hat{I}^{\epsilon_i})\,,
\end{align}
and we can combine ${\cal A}^{{\rm YM}\oplus{\rm Tr}(\phi^3)}(\hat{i}_\phi,A_{1L},\hat{I}_\phi)$ and ${\cal J}^{{\rm YM}\oplus{\rm Tr}(\phi^3)}(\hat{I}_\phi,A_{1R},\hat{j}_\phi,\kappa_{I\phi})$ together to obtain another analogue of \eqref{resA-resJ},
\begin{align}
&\sum_H\,{\cal A}^{\rm YM}(\hat{i},A_{1L},\hat{I})\,{1\over P_I^2}\,{\cal A}^{\rm YM}(\hat{I},A_{1R},\hat{j},B(k))\nn
=&-{\rm Res}\big|_{z=z_I}\,{{\cal J}_{n_1}^{{\rm YM}\oplus{\rm Tr}(\phi^3)}(\hat{i}_\phi,A,\hat{j}_\phi,\kappa_\phi)\over z}\,\times\,\epsilon_k\cdot{\cal J}_{n+3-n_1}^{\rm YM}(\hat{j},B(\kappa_I'),\hat{I}^{\epsilon_i})\nn
=&-{\rm Res}\big|_{z=z_I}\,{{\cal J}_{n_1}^{{\rm YM}\oplus{\rm Tr}(\phi^3)}(\hat{i}_\phi,A,\hat{j}_\phi,\kappa_\phi)\over z}\,\times\,\epsilon_k\cdot{\cal J}_{n+3-n_1}^{\rm YM}(\hat{j},B(\kappa'),\hat{i})\,,
\end{align}
From  the second line to the third line, we require the identity
\bea 
\epsilon_k\cdot{\cal J}_{n+3-n_1}^{\rm YM}(\hat{j},B(\kappa_I'),\hat{I}^{\epsilon_i})=\epsilon_k\cdot{\cal J}_{n+3-n_1}^{\rm YM}(\hat{j},B(\kappa'),\hat{i})\,.~~~\label{confusion-1}
\eea

However, the identity \eqref{confusion-1} is highly non-trivial since, based on general arguments similar to those in previous paragraphs, one cannot exclude the appearance of the contraction $\hat{k}_i\cdot \hat{k}_j$. Indeed, since the current is not gauge invariant, different gauge choices will lead to different expressions for the currents. Consequently, some of these will violate  \eqref{confusion-1}, while others will satisfy it. The one appearing in the 2-split is defined through the CHY-frame in \cite{Cao:2024gln,Cao:2024qpp} and it does satisfy \eqref{confusion-1}.

To see the gauge dependence of the current, let us compute the three point example. Using the rule of 3-vertex in Feynman gauge for color-ordered amplitudes 
\begin{align}
	V^{\mu_1\mu_2\mu_3}=\frac{i}{\sqrt{2}}\left[\eta^{\mu_1\mu_2}(p_1-p_2)^{\mu_3}+\eta^{\mu_2\mu_3}(p_2-p_3)^{\mu_1}+\eta^{\mu_3\mu_1}(p_3-p_1)^{\mu_2}\right]~~~\label{con-2-1}
\end{align}
we have 
\begin{align}
	{\cal A}_3=\frac{i}{\sqrt{2}}\left[\epsilon_1\cdot\epsilon_2~(p_1-p_2)\cdot\epsilon_3+\epsilon_2\cdot\epsilon_3~(p_2-p_3)\cdot\epsilon_1+\epsilon_1\cdot\epsilon_3~(p_3-p_1)\cdot\epsilon_2\right]~~~\label{con-2-2}
\end{align}
while using the 3-vertex in  the Gervais-Neveu(GN) gauge \cite{Gervais:1972tr},
\begin{align}
	V^{\mu_1\mu_2\mu_3}=i\sqrt{2}\left[\eta^{\mu_1\mu_2}(p_1-p_2)^{\mu_3}+\eta^{\mu_2\mu_3}(p_2-p_3)^{\mu_1}+\eta^{\mu_3\mu_1}(p_3-p_1)^{\mu_2}\right]~~~\label{con-3-1}
\end{align}
 it is:
\begin{align}
	{\cal A}_3=i\sqrt{2}\left[\epsilon_1\cdot\epsilon_2 ~p_1\cdot\epsilon_3+\epsilon_2\cdot\epsilon_3 ~p_2\cdot\epsilon_1+\epsilon_1\cdot\epsilon_3 ~p_3\cdot\epsilon_2\right]~~~\label{con-3-2}
\end{align}
It is apparent that for on-shell amplitudes, these expressions are equivalent under the  conditions $k_i^2=0, \epsilon_i\cdot k_i=0$. However, for the current, for example, $k_1^2\neq 0, k_1\cdot \epsilon_1\neq 0$, one can see that results \eqref{con-2-2} and \eqref{con-3-2} differ. 

One can check that for the current ${\cal J}(1,2,3,4)$ with $i=1,j=2,k=4$, if the GN gauge is adopted, the 
s-channel as
\begin{align}
	\frac{1}{s_{1,2}}V^{\mu_1\mu_2\mu}(1,2,I)V_{\nu\nu_1\nu_2}(-I,3,4)=-2(\epsilon_1\cdot\epsilon_2)(\epsilon_3\cdot\epsilon_4)(p_2\cdot p_3)+\ldots\,,
\end{align}
and the t-channel as
\begin{align}
	\frac{1}{s_{2,3}}V^{\mu_1\mu_2\mu}(4,1,I)V_{\nu\nu_1\nu_2}(-I,2,3)=-2(\epsilon_1\cdot\epsilon_4)(\epsilon_2\cdot\epsilon_3)(p_4\cdot p_3)+\ldots\,,
\end{align}
where the $\ldots$ are $p\cdot\epsilon$ terms. One can see that the $s_{12}$ does not appear\footnote{In fact, the result \eqref{con-3-2} is the  Eq.~(4.29) in \cite{Cao:2024qpp}}.

In general, in \cite{Cao:2024gln,Cao:2024qpp}, the current is defined with a specific choice of removing particular rows and columns during the construction of the CHY integrand, as well as employing the three scattering equations in the CHY measure. Using these specific choices, it has been shown in \cite{Naculich:2015zha} that for theories with at most three massive particles, all momentum contractions are of the form:
\bea k_c\cdot k_{\alpha},~~~~k_{\alpha}\cdot k_{\beta}~~~~\label{con-5}\eea
where $c$ are momenta of massive particles and $\alpha,\beta$ are the remaining massless momenta. For our current case, $c$ are $\hat{k}_i, \hat{k}_j,k_{k'_I}$ for $\epsilon_k\cdot{\cal J}_{n+3-n_1}^{\rm YM}(\hat{j},B(\kappa_I'),\hat{I}^{\epsilon_i})$, while three momenta are $\hat{k}_i, \hat{k}_j,k_{k'}$ for $\epsilon_k\cdot{\cal J}_{n+3-n_1}^{\rm YM}(\hat{j},B(\kappa'),\hat{i})$. Under the set of variables given in \eqref{con-5}, they are identical, thus demonstrating the identity \eqref{confusion-1} for this specific current.

The treatment for remaining three lines in \eqref{residue-splitYM} is analogous. In the rest of this subsection, we focus on the first case \eqref{kine-condi-splitYM2}, since the manipulation for other case is the same. We can directly get the analogue of \eqref{resA-resJ-2},
\begin{align}
	{\cal A}^{\rm YM}_n=\sum_I\,\sum_H\,{{\cal A}^{\rm YM}_L(z_I)\,{\cal A}^{\rm YM}_R(z_I)\over P_I^2}
=-\sum_I\,{\rm Res}\big|_{z=z_I}\,{{\cal B}^{\rm YM}_n(z)\over z}\,,~~\label{resA-resJ-YM}
\end{align}
where the function ${\cal B}^{\rm YM}_n(z)$ is defined as
\begin{align}
{\cal B}^{\rm YM}_n(z)=\epsilon_k\cdot{\cal J}^{\rm YM}_{n_1}(z)\,\times\,{\cal J}^{{\rm YM}\oplus{\rm Tr}(\phi^3)}_{n+3-n_1}(z)\,.
\end{align}

As in the previous ${\rm Tr}(\phi^3)$ case, to finish the proof, we need to consider the pole at $z=\infty$ of ${\cal B}^{\rm YM}_n(z)$. For ${\cal B}^{\rm YM}_n(z)$,  the background field method in \cite{Arkani-Hamed:2008bsc} applied to the two currents respectively tells us that: (a) For a YM amplitude/current with proper deformation the large  $z$ behavior is $z^{-1}$; (b) for a YM$\oplus{\rm Tr}(\phi^3)$ amplitude/current, with proper deformation the large $z$ behavior is $z^{1}$; (c) For both YM and YM$\oplus{\rm Tr}(\phi^3)$ cases, if two deformed momenta are non-adjacent in the color ordering, the power of $z$ can be reduced by $1$. 
Now, considering the two currents in ${\cal B}^{\rm YM}_n(z)$, and given that one current corresponds to non-adjacent particles in the color ordering, the large $z$ behavior is at most $z^{-1}\times z\times z^{-1}$. Therefore, we conclude that
\begin{align}
{\cal B}^{\rm YM}_n={\cal B}^{\rm YM}_n(0)=-\sum_I\,{\rm Res}\big|_{z=z_I}\,{{\cal B}^{\rm YM}_n(z)\over z}\,.~~\label{side2-YM}
\end{align}
Combining \eqref{resA-resJ-YM} and \eqref{side2-YM} together, we find ${\cal A}^{\rm YM}_n={\cal B}^{\rm YM}_n$, under the constraints
\eqref{kine-condi-splitphi}, \eqref{kine-condi-splitYM1}, \eqref{kine-condi-splitYM2}.

Before ending this part, let us 
remark that above naive analysis suggests the large $z$ behavior of ${\cal A}^{\rm YM}$ to be  $z^{-2}$ and ${\cal B}^{\rm YM}$ to be $z^{-1}$.
However,  since we have proved they are the same, it means that the actual large $z$ behavior of ${\cal B}^{\rm YM}$ is ${1\over z^2}$.

\subsection{GR amplitudes}
\label{subsec-GR-split}

In this subsection, we will prove two types of $2$-split of GR amplitudes. The First type is valid for both pure Einstein-Hilbert gravity and the extended version including dilatons and B-fields, while the second type is specific to  the extended theory. Given the similarity to the YM amplitude case, our discussion will be more concise.

The first split type is expressed as \cite{Cao:2024qpp}\footnote{It is worth to emphasize that for GR theory, there 
is no color ordering. Consequently, the sets $A$ and $B$ are equivalent, and interchanging $b$ with $a$ in \eqref{kine-condi-splitGR2} does not yield a new split pattern.}
\begin{align}
\label{split-GR1}
{\cal A}^{\rm GR}_{n}\,\xrightarrow[]{\eqref{kine-condi-splitphi}\eqref{kine-condi-splitGR1}\eqref{kine-condi-splitGR2}}\,\varepsilon_k\cdot{\cal J}^{\rm GR}_{n_1}(i,A(\kappa),j)\,\times\,{\cal J}^{{\rm GR}\oplus{\rm Tr}(\phi^3)}_{n+3-n_1}(j_\phi,B(\kappa'_\phi),i_\phi)\,,
\end{align}
with constraints on polarization tensors\footnote{In \cite{Cao:2024qpp} (see Eq.(4.30)), the conditions
$\epsilon_a\cdot \W\epsilon_b=\W\epsilon_a\cdot \epsilon_b=0$ are not imposed. But since all discussions are based on the CHY frame, where the Lorentz contractions $\epsilon_a\cdot \W\epsilon_b=\W\epsilon_a\cdot \epsilon_b=0$ do not appear at all, these extra conditions does not matter. }
\begin{align}
\{ k_a,\epsilon_a,\W\epsilon_a \}\cdot 	\{ k_b,\epsilon_b,\W\epsilon_b \}=0\,,~~~~{\rm for}~a\in A\,,~b\in B\,,
\label{kine-condi-splitGR1}
\end{align}
and
\begin{align}
\{ k_b,\epsilon_b,\W\epsilon_b \}\cdot \{\epsilon_c,\W\epsilon_c \}=0
	\,,~~~~{\rm with}~c=i,j,k\,.~~\label{kine-condi-splitGR2}
\end{align}
In the above, $\varepsilon^{\mu\nu}_q$ represents the polarization tensor carried by the external graviton $q$, and is decomposed as
$\varepsilon^{\mu\nu}_q\equiv\epsilon_q^\mu\W\epsilon_q^\nu$. The tensor current $\big({\cal J}^{\rm GR}_{n_1}(i,A,j,\kappa)\big)^{\mu\nu}$ carries an off-shell graviton $\kappa$, and is contracted with the polarization tensor $\varepsilon_k^{\mu\nu}$. 

Now we prove the $2$-split in \eqref{split-GR1} via the recursive technique. The lowest three-point GR amplitudes trivially factorize as ${\cal A}^{\rm GR}_3={\cal A}^{\rm GR}_3\times{\cal A}^{{\rm Tr}(\phi^3)}_3$. The factorizations of higher-point amplitudes are again determined by recursion. By deforming $k_i$ and $k_j$, one can immediately derive the analogue of \eqref{residue-splitYM},
\begin{align}
{\cal A}^{\rm GR}_{n}
=&\sum_{A_{1R}}\,\sum_H\,{\cal A}^{\rm GR}(\hat{i},A_{1L},\hat{I})\,{1\over P_I^2}\,{\cal A}^{\rm GR}(\hat{I},A_{1R}(k),B,\hat{j})\nn
 &+\sum_{A_{2L}}\,\sum_H\,{\cal A}^{\rm GR}(B(k),\hat{i},A_{2L},\hat{I})\,{1\over P_I^2}\,{\cal A}^{\rm GR}(\hat{I},A_{2R},\hat{j})\nn
 &+\sum_{B_{3R}}\,\sum_H\,{\cal A}^{\rm GR}(B_{3L},\hat{i},\hat{I})\,{1\over P_I^2}\,{\cal A}^{\rm GR}(\hat{I},A,\hat{j},B_{3R}(k))\nn
 &+\sum_{B_{4L}}\,\sum_H\,{\cal A}^{\rm GR}(B_{4L},\hat{i},A,\hat{I})\,{1\over P_I^2}\,{\cal A}^{\rm YM}(\hat{j},B_{4R},\hat{I})\,.
 \label{residue-splitGR1}
\end{align}
Here, we have used the fact that the large $z$ behavior of ${\cal A}^{\rm GR}_n(z)$ is  $z^{-2}$ as shown in \cite{Arkani-Hamed:2008bsc}. Repeating the process from the previous YM case, with polarization vectors replaced by polarization tensors, the analogue of \eqref{resA-resJ-YM} is found to be
\begin{align}
{\cal A}^{\rm GR}_n=\sum_I\,\sum_H\,{{\cal A}^{\rm GR}_L(z_I)\,{\cal A}^{\rm GR}_R(z_I)\over P_I^2}
=-\sum_I\,{\rm Res}\big|_{z=z_I}\,{{\cal B}^{\rm GR}_n(z)\over z}\,,~~\label{resA-resJ-GR1}
\end{align}
where
\begin{align}
{\cal B}^{\rm GR}_n(z)=\epsilon_k\cdot{\cal J}^{\rm GR}_{n_1}(z)\,\times\,{\cal J}^{{\rm GR}\oplus{\rm Tr}(\phi^3)}_{n+3-n_1}(z)\,.
\end{align}
We want to emphasize that to arrive \eqref{resA-resJ-GR1}, two similar technical points need to be addressed. Firstly, the polarization vectors of $\hat{I}$ must satisfy the split condition \eqref{kine-condi-splitGR2}. A non-rigorous argument, analogous to the YM case, can be constructed, i.e., by dividing  the full kinematic and polarization space into two orthogonal complementary subspaces, namely ${\cal S}={\cal S}_1\oplus{\cal S}_2$. Secondly, identities like \eqref{confusion-1}, which are crucial for the proof, are required. As in the YM case, the corresponding currents are special ones, i.e., defined by the CHY integrand with a specific gauge choice, as discussed in \cite{Naculich:2015zha}.

For large $z$ behavior of  ${\cal B}^{\rm GR}_n(z)$, we have  $\epsilon_k\cdot{\cal J}^{\rm GR}_{n_1}(z)\sim z^{-2}$, while ${\cal J}^{{\rm GR}\oplus{\rm Tr}(\phi^3)}_{n+3-n_1}(z)\sim z^1$. The reason for the later  is as follows: when a GR$\oplus{\rm Tr}(\phi^3)$ amplitude/current  contains two on-shell external scalars $i$ and $j$, the large $z$ behaves is $z^2$ \cite{Arkani-Hamed:2008bsc}. If adding a third scalar along the line connecting $i,j$ involves replacing one graviton-scalar-scalar cubic vertex with a scalar-scalar-scalar cubic vertex, which reduces the large $z$ behavior by at least a factor ${1\over z}$, i.e., $z^2\times {1\over z}=z$. Putting all together, we get a ${1\over z}$ behavior for  the ${\cal B}^{\rm GR}_n(z)$ side, so we  have
\begin{align}
{\cal B}^{\rm GR}_n={\cal B}^{\rm GR}_n(0)=-\sum_I\,{\rm Res}\big|_{z=z_I}\,{{\cal B}^{\rm GR}_n(z)\over z}\,.~~\label{side2-GR1}
\end{align}
Substituting \eqref{resA-resJ-GR1} and \eqref{side2-GR1} into \eqref{resA-resJ-GR1}, we obtain ${\cal A}^{\rm GR}_n={\cal B}^{\rm GR}_n$, thus the proof is finished.

The second type of $2$-split is given as \cite{Cao:2024qpp}
\begin{align}
{\cal A}^{\rm GR}_{n}\,\xrightarrow[]{\eqref{kine-condi-splitphi}\eqref{kine-condi-splitGR1}\eqref{kine-condi-splitGR-EY}}\,\epsilon_k\cdot{\cal J}^{\rm EYM}_{n_1}(i_g,A,j_g,\kappa_g)\,\times\,\W\epsilon_k\cdot{\cal J}^{\rm EYM}_{n+3-n_1}(j_g,B(\kappa'_g),i_g)\,,~~~~\label{split-GR-EY}
\end{align}
with new constraints on polarization tensors $\varepsilon_i$, $\varepsilon_j$ and $\varepsilon_k$,
\begin{align}
\epsilon_c\cdot \{k_b, \epsilon_b,\W\epsilon_b\}=0\,,~~~~\W\epsilon_c\cdot \{k_a, \epsilon_a,\W\epsilon_a\}=0\,,
~~~~{\rm with}~c=i,j,k\,.
\label{kine-condi-splitGR-EY}
\end{align}
In condition \eqref{kine-condi-splitGR-EY}, $\W\epsilon_c$ can never be chosen as $\epsilon_c$. Consequently, the corresponding $2$-split \eqref{split-GR-EY} only make sense for extended gravity with general two-index polarization tensors. In \eqref{split-GR-EY}, each resulting current is a vector current carrying three external gluons $i_g$, $j_g$ and $\kappa_g(\kappa'_g)$.

The $2$-split in \eqref{split-GR-EY} can also be proven using the recursive approach. In this case, the factorization of lowest-point amplitude is indicated by the well-known double copy construction, ${\cal A}^{\rm GR}_3={\cal A}^{\rm YM}_3\times{\cal A}^{\rm YM}_3$. To perform the recursion, we again deform $k_i$ and $k_j$ to obtain \eqref{residue-splitGR1}. we then encounter a new situation when reducing the residues of the full amplitude to the residues of one of the two currents. Taking the first line of \eqref{residue-splitGR1} as the example, the sub-amplitude ${\cal A}^{\rm GR}(\hat{i},A_{1L},\hat{I})$ is linear in $\epsilon_i$ and $\W\epsilon_i$ lying in ${\cal S}_1$ and ${\cal S}_2$ respectively. The polarization $\W\epsilon_i$ should be contracted with $\W\epsilon_I^H$; otherwise, ${\cal A}^{\rm GR}(\hat{i},A_{1L},\hat{I})$ will be annihilated. Thus, we can use \eqref{polar-metri} to absorb $\W\epsilon_i$ into ${\cal A}^{\rm GR}(\hat{I}^{\W\epsilon_i},A_{1R},\hat{j},B(k))$. After this step, the GR amplitude ${\cal A}^{\rm GR}(\hat{i},A_{1L},\hat{I})$ is converted to the EYM one ${\cal A}^{{\rm EYM}}(\hat{i}_g,A_{1L},\hat{I}_g)$, which carries two external gluons $\hat{i}_g$ and $\hat{I}_g$. All kinematic variables carried by ${\cal A}^{{\rm EYM}}(\hat{i}_g,A_{1L},\hat{I}_g)$ lie in ${\cal S}_1$, except $\hat{k}_i$. Then we can use momentum conservation to remove $\hat{k}_i$, and find that the effective $\epsilon^H_I$ should lie in ${\cal S}_1$. This observation forces $\bar{\epsilon}^H_I$ of ${\cal A}^{\rm GR}(\hat{I}^{\W\epsilon_i},A_{1R},\hat{j},B(k))$ to be in ${\cal S}_1$. Thus, in the resulting sub-amplitude ${\cal A}^{\rm GR}(\hat{I}^{\W\epsilon_i},A_{1R},\hat{j},B(k))$, the external graviton $\hat{I}$ carries two polarization vectors $\bar{\epsilon}^H_I$ and $\W\epsilon_i$, lying in ${\cal S}_1$ and ${\cal S}_2$ respectively. This configuration exactly satisfies the condition \eqref{kine-condi-splitGR-EY}, allowing ${\cal A}^{\rm GR}(\hat{I}^{\W\epsilon_i},A_{1R},\hat{j},B(k))$ to factorize as in \eqref{split-GR-EY},
\begin{align}
{\cal A}^{\rm GR}(\hat{I}^{\W\epsilon_i},A_{1R},\hat{j},B(k))\xrightarrow[\eqref{kine-condi-splitGR1}\eqref{kine-condi-splitGR-EY}]{\eqref{kine-condi-splitphi}}
\,\epsilon_k\cdot{\cal J}^{\rm EYM}(\hat{I}_g,A_{1R},\hat{j}_g,\kappa_{g})\times\W\epsilon_k\cdot{\cal J}^{\rm EYM}(\hat{j}_g,B(\kappa'_{I,g}),\hat{I}^{\W\epsilon_i}_g)\,.
\end{align}
The above $2$-split allows us to reform each term in the first line of \eqref{residue-splitGR1} as
\begin{align}
 &\sum_H\,{\cal A}^{\rm GR}(\hat{i},A_{1L},\hat{I})\,{1\over P_I^2}\,{\cal A}^{\rm GR}(\hat{I},A_{1R},\hat{j},B(k))\nn
=&-{\rm Res}\big|_{z=z_I}\,
{\epsilon_k\cdot{\cal J}^{\rm EYM}_{n_1}(\hat{i}_g,A,\hat{j}_g,\kappa_g)\over z}\,\times\,\W\epsilon_k\cdot{\cal J}^{\rm EYM}_{n+3-n_1}(\hat{j}_g,B(\kappa'_g),\hat{i}_g)\,,
\end{align}
by gluing ${\cal A}^{{\rm EYM}}(\hat{i}_g,A_{1L},\hat{I}_g)$ and $\epsilon_k\cdot{\cal J}^{\rm EYM}(\hat{I}_g,A_{1R}\hat{j}_g,\kappa_g)$
together. In the above, we have used
\begin{align}
\W\epsilon_k\cdot{\cal J}^{\rm EYM}_{n+3-n_1}(\hat{j}_g,B(\kappa'_{I,g}),\hat{I}^{\W\epsilon_i}_g)=\W\epsilon_k\cdot{\cal J}^{\rm EYM}_{n+3-n_1}(\hat{j}_g,B(\kappa'_g),\hat{i}_g)\,,
\end{align}
for the same reason as in the YM case.

Based on the above treatment, we obtain the following conclusion for the recursive part
\begin{align}
\sum_I\,\sum_H\,{{\cal A}^{\rm GR}_L(z_I)\,{\cal A}^{\rm GR}_R(z_I)\over P_I^2}=-\sum_I\,{\rm Res}\big|_{z=z_I}\,{{\cal D}^{\rm GR}_n(z)\over z}\,,~~\label{resA-resJ-GR2}
\end{align}
where
\begin{align}
{\cal D}^{\rm GR}_n(z)=\epsilon_k\cdot{\cal J}^{\rm EYM}_{n_1}(z)\,\times\,\W \epsilon_k\cdot{\cal J}^{\rm EYM}_{n+3-n_1}(z)\,.
\end{align}
Now we need to deal with the boundary contributions of the ${\cal D}^{\rm GR}_n(z)$. As shown in \cite{Arkani-Hamed:2008bsc}, when an Einstein-Maxwell amplitude/current carries two external on-shell photons $i$ and $j$, the upper limit of large $z$ behavior under the deformation of $k_i$ and $k_j$ is $z^0$. This result also holds for the EYM amplitude/current with two external on-shell gluons $i_g$ and $j_g$, due to the absence of self-interaction of gluons. However, for current case,  each EYM current in ${\cal D}^{\rm GR}_n(z)$ has three external gluons rather than two. Turning the graviton $k^*_h$ into a gluon $k^*_g$ converts the ${\cal O}(z^2)$
vertex $g-g-h$ to the ${\cal O}(z)$ one $g-g-g$, implying that the large $z$ behavior of each current is $z^{-1}$. With ${\cal D}^{\rm GR}_n(z)$ vanishing at the infinity,  we find
\begin{align}
{\cal D}^{\rm GR}_n={\cal D}^{\rm GR}_n(0)=-\sum_I\,{\rm Res}\big|_{z=z_I}\,{{\cal D}^{\rm GR}_n(z)\over z}\,.~~\label{side2-GR2}
\end{align}
Combining \eqref{resA-resJ-GR2} and \eqref{side2-GR2} together, we get ${\cal A}^{\rm GR}_n={\cal D}^{\rm GR}_n$, which states the $2$-split in \eqref{split-GR-EY}.

\subsection{NLSM amplitudes}
\label{subsec-NLSM-split}

The NLSM amplitude with even $n$ has the following $2$-split \cite{Cao:2024qpp}
\begin{align}
{\cal A}^{\rm NLSM}_{n}(1,\cdots,n)\,\xrightarrow[]{\eqref{kine-condi-splitphi}}\,{\cal J}^{\rm NLSM}_{n_1}(\kappa,i,A,j)\,\times\,{\cal J}^{{\rm NLSM}\oplus{\rm Tr}(\phi^3)}_{n+3-n_1}(j_\phi,B(\kappa'_\phi),i_\phi)\,,
\label{split-NLSM1}
\end{align}
where the number of elements in $A$ is odd and the number of elements in $B$ is even, and
\begin{align}
{\cal A}^{\rm NLSM}_{n}(1,\cdots,n)\,\xrightarrow[]{\eqref{kine-condi-splitphi}}\,{\cal J}^{{\rm NLSM}\oplus{\rm Tr}(\phi^3)}_{n_1}(\kappa_\phi,i_\phi,A,j_\phi)\,\times\,{\cal J}^{\rm NLSM}_{n+3-n_1}(j,B(\kappa'),i)\,,~~~~\label{split-NLSM2}
\end{align}
for the case with exchanged odd and even numbers of elements in $A$ and $B$. In the above expression, the subscript $p$ stands for pion, while $\phi$ stands for scalar of ${\rm Tr}(\phi^3)$ theory.

In this instance, we are not able to prove the $2$-split in \eqref{split-NLSM1} and \eqref{split-NLSM2} using the BCFW recursion relation. The primary obstacle lies in our lack of understanding of how to handle the pole at infinity. However, for poles at finite locations, we can perform analogous manipulations. 

The lowest $4$-point NLSM amplitudes again trivially factorize as
\begin{align}
{\cal A}^{\rm NLSM}_4={\cal A}^{\rm NLSM}_4\,\times\,{\cal A}^{{\rm Tr}(\phi^3)}_3\,.
\end{align}
Meanwhile, by deforming $k_i$ and $k_j$, and utilizing hidden zeros, one can also find the analogue of \eqref{residue-splitphi},
\begin{align}
{\cal A}^{{\rm NLSM}}(1,\cdots,n)|_{finite}
=&\sum_{A_{1R}}\,{\cal A}^{{\rm NLSM}}(\hat{i},A_{1L},\hat{I})\,{1\over P_I^2}\,{\cal A}^{\phi^3}(\hat{I},A_{1R},j,B(k))\nn
& + \sum_{A_{2L}}\,{\cal A}^{{\rm NLSM}}(B(k),\hat{i},A_{2L},\hat{I})\,{1\over P_I^2}\,{\cal A}^{\phi^3}(\hat{I},A_{2R},\hat{j})\nn
& + \sum_{B_{3R}}\,{\cal A}^{{\rm NLSM}}(B_{3L},\hat{i},\hat{I})\,{1\over P_I^2}\,{\cal A}^{\phi^3}(\hat{I},A,\hat{j},B_{3R}(k))\nn
& + \sum_{B_{4L}}\, {\cal A}^{{\rm NLSM}}(B_{4L}(k),\hat{i},A,\hat{I})\,{1\over P_I^2}\,{\cal A}^{\phi^3}(\hat{j},B_{4R},\hat{I})\,.
\label{residue-NLSM}
\end{align}
Here, we  emphasize that the right-hand side includes only contributions from finite poles. Then, by employing the $2$-split of lower-point amplitudes, one can get the analogue of \eqref{resA-resJ-2},
\begin{align}
\sum_I\,{{\cal A}^{\rm NLSM}_L(z_I)\,{\cal A}^{\rm NLSM}_R(z_I)\over P_I^2}
=-\sum_I\,{\rm Res}\big|_{z=z_I}\,{{\cal B}^{\rm NLSM}_n(z)\over z}\,,~~\label{resA-resJ-NLSM}
\end{align}
through an extremely similar argument, where
\begin{align}
{\cal B}^{\rm NLSM}_n(z)={\cal J}^{\rm NLSM}_{n_1}(z)\,\times\,{\cal J}^{\rm NLSM}_{n+3-n_1}(z)\,.
\end{align}

However, as discussed in previous subsections, the proof can be completed along this line if and only if ${\cal A}^{\rm NLSM}_n(z)$ and ${\cal B}^{\rm NLSM}_n(z)$ have the same residue for pole at the infinity, which is not zero. We can suppress the large $z$ behavior of ${\cal A}^{\rm NLSM}_n(z)$ by using  the modified BCFW recursion relation as done in \eqref{2.16}, i.e., considering the contour integral:
\begin{align}
\frac{1}{2\pi i}\oint\,{dz\over z}\,{{\cal A}^{\rm NLSM}(z)\over \prod_{r=1}^{a+1}\,(z_r-z)}\,,
\end{align}
where $z_r$ are $a+1$ roots of
\begin{align}
{\cal A}^{\rm NLSM}_n(z)={N(m+a)\over D(m)}\,.
\end{align}
In the above modified integral, the integrand vanishes at infinity, thus we get
\begin{align}
{{\cal A}^{\rm NLSM}_n\over\prod_{r=1}^{a+1}\,z_r}=\sum_I\,{1\over \prod_{r=1}^{a+1}\,(z_r-z_I)}\,{{\cal A}^{\rm NLSM}_L(z_I)\,{\cal A}^{\rm NLSM}_R(z_I)\over P_I^2}\,.
\end{align}
By utilizing the previous technique, one can extend the above relation to
\begin{align}
{{\cal A}^{\rm NLSM}_n\over\prod_{r=1}^{a+1}\,z_r}=-\sum_I\,{{\rm Res}\big|_{z=z_I}\,{\cal B}^{\rm NLSM}_n(z)\over z_I\, \prod_{r=1}^{a+1}\,(z_r-z_I)}\,.~~~\label{resA-resJ-NLSM2}
\end{align}
then the problem is shifted to the proof of
\begin{align}
{{\cal B}^{\rm NLSM}_n\over\prod_{r=1}^{a+1}\,z_r}=-\sum_I\,{{\rm Res}\big|_{z=z_I}\,{\cal B}^{\rm NLSM}_n(z)\over z_I\, \prod_{r=1}^{a+1}\,(z_r-z_I)}\,,
\end{align}
or equivalently,
\begin{align}
{{\cal B}^{\rm NLSM}_n\over\prod_{r=1}^{a+1}\,z_r}={1\over 2\pi i}\oint\,{dz\over z}\,{{\cal B}^{\rm NLSM}_n(z)\over \prod_{r=1}^{a+1}\,(z_r-z)}\,.
\end{align}
The above equation is equivalent to the statement that  $z_r$ is also a zero of ${\cal B}^{\rm NLSM}_n$, which is far from obvious. 

In summary, using the BCFW recursion relation, we can only verify the  $2$-split \eqref{split-NLSM1} and \eqref{split-NLSM2} to a certain extent, specifically that they share the same residues at finite poles.
 
\section{\label{sec:con}Conclusion}

In this note, we have investigated the recently discovered hidden zeros in tree-level amplitudes and their associated 2-split structures, using the BCFW on-shell recursion. To address amplitudes that do not exhibit vanishing boundary terms—such as those in the NLSM—we proposed a modified contour integral. This modified contour framework allows one to rigorously prove the existence of hidden zeros in a wide class of theories, including $\mathrm{Tr}(\phi^3)$, YM, NLSM, SG, DBI, and GR. 

Furthermore, by leveraging the simple factorizations of low-point amplitudes and the hidden zeros structure, we recursively established the 2-split forms of $\mathrm{Tr}(\phi^3)$, YM and GR amplitudes, and verified the residues at physical poles for the 2-split of NLSM amplitudes. A crucial and nontrivial issue in this recursive construction is the definition of the resulting currents. While these currents are uniquely determined by the CHY formalism, aspects such as gauge choices and their associated implications warrant further clarification in future work.

Interestingly, all currently known hidden zeros correspond to theories with known CHY (or stringy) representations. This raises a natural question: for a theory lacking a CHY/stringy formulation, how can one determine whether there are hidden zeros for amplitudes? One can also ask are there other zero conditions different from the one found in current literature? For these questions,  BCFW recursion relations may offer an approach to address them. By computing low-point amplitudes directly, one can find the corresponding zero conditions. If these conditions can be persisted to higher points by recursion, it is possible such hidden zero exists for general amplitudes from the point of view of recursion relation. Thus, the problem of identifying the origin of hidden zeros is effectively reduced to analyzing a limited set of low-point amplitudes.

Recently, hidden zero for one-loop integral has been studied in \cite{Backus:2025hpn}, it will be interesting to use 
recursion relation to study hidden zero for more general theories at the loop level. We can also consider other properties, such as Adler zero, from this aspect. 
 
\section*{Acknowledgments}

We would like to thank Qu Cao for  valuable discussions. This work was supported in part by the Grants No.NSFC-12035008 and No.NSFC-12475122, by the Guangdong Major Project of Basic and Applied Basic Research No.2020B0301030008, and by the Fundamental Research Funds for the Central Universities (Grant No. 010-63253121); and under Grant No.11935013, No.11947301, No.12047502 (Peng Huanwu Center), No.12247103, and No.U2230402.

\bibliography{reference}
\bibliographystyle{JHEP}

\end{document}